\documentclass[10pt]{article}
\pdfoutput=1
\usepackage{chicago,epic,graphpap,graphics,float,tabularx,paralist,longtable,fancyhdr,fancyvrb,bbm}
\usepackage{amsmath}
\usepackage{amstext}
\usepackage{amsbsy}
\usepackage{amsthm}
\usepackage{amsfonts}
\usepackage{amssymb}
\usepackage{amscd}
\usepackage{eucal}
\def\d{\delta}

\def\th{\theta}
\def\g{\gamma}

\def\empty{\varnothing}
\def\as{\quad\text{{\rm a.s.}}}

\def\1{{\mathbbm 1}}
\def\E{{\mathbb E}}

\def\eqdef{\triangleq}

\def\sumi{\sum_{i=1}^n}
\def\s{\sigma}
\def\F{{\cal F}}

\def\X{\widetilde{X}}

\def\ito{It\^o}

\def\l{\lambda}

\begin{document}

\centerline{\LARGE\bf A rank-based approach to Zipf's law}
\vspace{10pt} \centerline{\large Ricardo T. Fernholz\footnote{Robert Day School of Economics and Finance, Claremont McKenna College, 500 E. Ninth St., Claremont, CA 91711, rfernholz@cmc.edu.} \hspace{10pt} \hspace{10pt} Robert Fernholz\footnote{INTECH, One Palmer Square, Princeton, NJ 08542.  bob@bobfernholz.com. The authors thank the members of the INTECH/Princeton SPT seminar for their comments and suggestions regarding this research.}} 
\vspace{5pt}
\centerline{\today}
\vspace{10pt}
\begin{abstract}
An Atlas model is a rank-based system of continuous semimartingales for which the steady-state values of the processes follow a power law, or Pareto distribution. For a power law, the log-log plot of these steady-state values versus rank is a straight line. Zipf's law is a power law for which the slope of this line is $-1$.  In this note, rank-based conditions are found under which an Atlas model will follow Zipf's law. An advantage of this rank-based approach is that it provides information about the dynamics of systems that result in Zipf's law.
\end{abstract}
\vspace{5pt}

\noindent{\large\bf Introduction}
\vspace{8pt}

A family of random variables follows a {\em power law,} or {\em Pareto distribution,} if the log-log plot of their values versus rank forms (approximately) a straight line. The random variables follow {\em Zipf's law} if the slope of this line is $-1$. \citeN{Newman:2006} and \citeN{Gabaix:2009} both present surveys of many different power laws observed in the real world. A characterization of conditions that result in Zipf's law for the population of cities is presented in \citeN{Gabaix:1999}, and this characterization is based on the idea that under a stable distribution the expected change in the population of each individual city is zero, at least when the city is away from a reflecting lower barrier. 

In the setting of Atlas models and other systems of rank-based continuous semimartingales (see \citeN{F:2002}), we examine the conditions that give rise to Zipf's law and consider several generalizations that are common in the real world. We shall find that this new setting is natural for an understanding of Zipf's law and provides insight into the dynamics involved.

\vspace{10pt}
\noindent{\large\bf Atlas models}
\vspace{8pt}

An Atlas model is a family of positive continuous semimartingales $X_1,\ldots,X_n$, with $n\ge2$, that satisfy
\begin{equation}\label{0}
d\log X_i(t) = \big(\g-g+ng\1_{\{r_t(i)=n\}}\big)dt+\s\,dW_i(t),
\end{equation}
for $i=1,\ldots,n$, where $\g$ is a constant,  $g$ and $\s$ are positive constants, $r_t(i)$ is the rank of $X_i(t)$ (with ties resolved lexicographically), and $W_1,\ldots,W_n$ is an $n$-dimensional Brownian motion with the Brownian filtration $\F_t$ (see \citeN{F:2002}, \shortciteN{BFK:2005}, and \citeN{FK:2009}). The processes $X_i$ might represent, for example, the wealth of households, the capitalizations of companies, or the population of cities. Let  $X_{(1)}\ge\cdots\ge X_{(n)}$ represent the ranked processes $X_1,\ldots,X_n$, so $X_{(r_t(i))}(t)=X_i(t)$.  We can define the {\em total value process} $X$ by
\[
X(t)\eqdef X_i(t)+\cdots+X_n(t),
\]
and the {\em weight processes} $\th_i$ and the {\em ranked weight processes} $\th_{(k)}$  by 
\[
\th_i(t) \eqdef X_i(t)/X(t)\quad\text{ and }\quad \th_{(k)}(t)\eqdef X_{(k)}(t)/X(t),\quad\text{ for }\quad i,k=1,\ldots,n.
\]

The term $ng\1_{\{r_t(i)=n\}}$ in \eqref{0} is a device that stabilizes the model by driving the ``Atlas''  process $X_{(n)}$ upward at a rate that counteracts the general downward drift of $-g$. The Atlas process can be thought of as the ``birth and death'' of processes in the lowest ranks, as is common in the firm size and income distribution literatures in economics (see \citeN{Luttmer:2011} and \shortciteN{Gabaix/Lasry/Lions/Moll:2015}). It can also be thought of as a proxy for a system that extends infinitely downward, as in the infinite models of \citeN{palpit:2008} and \citeN{chapal:2009}.

The parameter $\g$ in \eqref{0} represents the growth rate of the entire system. Since here we are interested in relative behavior under steady-state conditions, we can assume that $\g=0$, and we shall do so from here on. In this case, definition \eqref{0} reduces to
\begin{equation}\label{1}
d\log X_i(t) = \big(-g+ng\1_{\{r_t(i)=n\}}\big)dt+\s\,dW_i(t),
\end{equation}
for $i=1,\ldots,n$. With this defining equation, the asymptotic growth rate of each of the $X_i$ will be zero, so 
\begin{equation*}
\lim_{t\to\infty}t^{-1}\log X_i (t)=0,\as,
\end{equation*}
for $i=1,\ldots,n$ (see, e.g.,  \citeN{F:2002}, \shortciteN{BFK:2005}, or \citeN{FK:2009}). 

By \ito's rule, it follows from \eqref{1} that
\begin{equation}\label{1.1}
dX_i(t)=\bigg(-g+\frac{\s^2}{2}+ng\1_{\{r_t(i)=n\}}\bigg)X_i(t)\,dt+\s X_i(t)\,dW_i(t),\as,
\end{equation}
for $i=1,\ldots,n$. From this we see that while the asymptotic growth rate of the system \eqref{1} is zero, the local behavior of the individual processes $X_i$ is more complicated.

For the model \eqref{1}, when the gap processes $\log X_{(k)}-\log X_{(k+1)}$ are in their steady-state distribution, these gaps  are exponentially distributed with
\begin{equation}\label{1.2}
\E\big[\log X_{(k)}(t)-\log X_{(k+1)}(t)\big]=\E\big[\log \th_{(k)}(t)-\log \th_{(k+1)}(t)\big]=\frac{\s^2}{2kg},
\end{equation}
for $k=1,\ldots,n-1$   (see \shortciteN{BFK:2005}).   It follows from \eqref{1.2} that 
\begin{equation}\label{1.21}
\frac{\E\big[\log\th_{(k)}(t)-\log\th_{(k+1)}(t)\big]}{\log k-\log(k+1)}=\frac{\s^2}{2kg(\log k -\log(k+1))}\cong-\frac{\s^2}{2g},\as,
\end{equation}
for $k=1,\ldots,n-1$, and the log-log plot of $\th_{(1)}(t),\ldots,\th_{(n)}(t)$ versus rank, which is  called the {\em distribution curve} of the model, is  approximately  a straight line with  (log-log) slope  $-\s^2/2g$. Therefore, we have approximately
\begin{equation}\label{1.3}
\th_{(k)}(t)\propto  k^{-\s^2/2g},\as,
\end{equation}
for $k=1,\ldots,n$, and we say that the Atlas model \eqref{1} has a {\em Pareto distribution} with {\em parameter} $\l$, where
\[
\l=\frac{\s^2}{2g}.
\]

\vspace{5pt}
\noindent{\large\bf Zipfian Atlas models}
\vspace{8pt}

Zipf's law is a Pareto distribution with parameter $\l=1$, so the ranked weights in \eqref{1.3}  will satisfy, approximately,
\begin{equation}\label{1.4}
\th_{(k)}(t)\propto  k^{-1},\as,
\end{equation}
for $k=1,\ldots,n$, and the distribution curve will be a straight line with slope $-1$ (see \citeN{Gabaix:1999} or \citeN{Gabaix:2009}). \citeN{Gabaix:1999} constructs a simple economic model that yields a Pareto distribution for city size and then normalizes city populations so that the expected population of each city is constant, at least when a city is away from a lower reflecting barrier. This requirement of constant expected population results in Zipf's law.

For an Atlas model, when $r_t(i)< n$, equation \eqref{1.1} becomes
\begin{equation}\label{2}
dX_i(t)=\bigg(-g+\frac{\s^2}{2}\bigg)X_i(t)\,dt+\s X_i(t)\,dW_i(t),\as,
\end{equation}
and
\begin{equation}\label{2.1}
\E\big[dX_i(t)\big|\F_t, r_t(i)<n\big]=0,\as,
\end{equation}
if
\begin{equation}\label{2.2}
\bigg(\frac{\s^2}{2g}-1\bigg)X_i(t)=0, \as
\end{equation}
Since $X_i(t)>0$, a.s., the necessary and sufficient condition for this to hold is that $\s^2/2g=1$. This condition is equivalent to the requirement that \eqref{2} be a martingale for $r_t(i)<n$ (see also \citeN{Bruggeman:2016}, Section~3.6).

\vspace{10pt}
\noindent{\bf Definition 1.} An Atlas model of the form \eqref{1} is {\em Zipfian} if $\s^2/2g=1$.
\vspace{10pt}

For a Zipfian Atlas model,
\[
\l=\frac{\s^2}{2g}=1,
\]
so it follows from \eqref{1.3} that \eqref{1.4} holds, which is exactly Zipf's law.
We shall see in the next section that although a Zipfian Atlas model is distributed according to Zipf's law, a model that follows Zipf's law is not necessarily Zipfian.

\vspace{10pt}
\noindent{\large\bf Weakly Zipfian Atlas models}
\vspace{8pt}

Empirically, for observed Zipf-like distributions it is not unusual for the distribution curve to be concave with the slope of the tangent flatter than $-1$ for the higher ranks and steeper than $-1$ for the lower ranks. Figure~I in \citeN{Gabaix:1999}, Figure~5.1 in \citeN{F:2002}, and Figure~11 in \citeN{Fernholz/Koch:2016} document this tendency, respectively, for the population of U.S.\ cities, the total market capitalizations of U.S.\ stocks, and the assets of U.S. bank holding-companies after the 1990s. The changing slopes in these real-world phenomena could be the result of variances $\s^2_k$ that increase with rank, as is conjectured for city size by \citeN{Gabaix:2009}, and is documented for stock capitalizations in Figure~5.5 of \citeN{F:2002}.

In order to study the case of increasing variances, let us consider a generalized Atlas model with variances that depend on rank. For $n\ge2$, let
\begin{equation}\label{8}
d\log X_i(t) = \big(-g+ng\1_{\{r_t(i)=n\}}\big)dt+\s_{r_t(i)}dW_i(t),
\end{equation}
for $i=1,\ldots,n$, where $g$ and $\s_1,\ldots,\s_n$ are positive constants, $r_t(i)$ is the rank of $X_i(t)$, and $W_1,\ldots,W_n$ is an $n$-dimensional Brownian motion (see, e.g., \citeN{F:2002} or \shortciteN{BFK:2005}). Here all the ranks share a common reversion rate $g$, but each rank $k$ has its own variance rate $\s^2_k$. \shortciteN{BFK:2005} show that for a system of this form, if the gap processes $\log X_{(k)}-\log X_{(k+1)}$ are in their steady-state distribution, then
\[
\E\big[\log \th_{(k)}(t)-\log \th_{(k+1)}(t)\big]=\frac{\s_k^2+\s_{k+1}^2}{4kg},
\]
for $k=1,\ldots,n-1$, so the tangent to the distribution curve between rank $k$ and rank $k+1$ has log-log slope of 
\begin{equation}\label{8.0}
\frac{\E\big[\log\th_{(k)}(t)-\log\th_{(k+1)}(t)\big]}{\log k-\log(k+1)}=\frac{\s_k^2+\s_{k+1}^2}{4kg(\log k -\log(k+1))}\cong-\frac{\s_k^2+\s_{k+1}^2}{4g}.
\end{equation}
 Let us note that this slope is consistent with the slope \eqref{1.21} for the standard Atlas model \eqref{1}. 
 
From \eqref{8.0} we can construct an example of a generalized Atlas model for which Zipf's law holds, but the model is not Zipfian. For an even number $n$ and $g>0$, let
 \[
 \s^2_{2j}=g\quad\text{ and }\quad\s^2_{2j+1}=3g,
 \]
 for $j=1,\ldots,n/2$. For these values, by \eqref{8.0},
 \[
 \frac{\E\big[\log\th_{(k)}(t)-\log\th_{(k+1)}(t)\big]}{\log k-\log(k+1)}\cong-\frac{\s_k^2+\s_{k+1}^2}{4g}=-1,
 \]
 for $k=1,\ldots,n-1$, so  the log-log slope of the tangent to the distribution curve is $-1$ for all ranks, which means that Zipf's law holds. However, we see from \eqref{8.1} that $ \s^2_{r_t(i)}/2g\ne1$ for any  $X_i$. Hence, Zipf's law holds for a Zipfian Atlas model, but a generalized Atlas model for which Zipf's law holds need not be Zipfian.
 
For a generalized Atlas model  \eqref{8}, equation \eqref{1.1} becomes 
\begin{equation}\label{8.1}
dX_i(t)=\bigg(-g+\frac{\s_{r_t(i)}^2}{2}+ng\1_{\{r_t(i)=n\}}\bigg)X_i(t)\,dt+\s_{r_t(i)} X_i(t)\,dW_i(t),\as,
\end{equation}
 for $i=1,\ldots,n$. Let us assume that the gap processes $\log X_{(k)}-\log X_{(k+1)}$ are in their steady-state distribution.  For variable $\s^2_k$, we cannot expect that $\s^2_{r_t(i)}/2g=1$ for all  $i$ with $r_t(i)<n$, so this model cannot be Zipfian. Instead, a more general definition is needed, so let us consider the {\em adjusted total value process}  $\X$ defined by
\begin{equation}\label{3.0}
d\X(t)\eqdef  dX(t) - ngX_{(n)}(t)\,dt.
 \end{equation}
We would like to impose conditions such that 
\begin{equation}\label{3}
\E\big[d\X(t)\big|\F_t\big]=0,\as,
\end{equation}
which  is a natural generalization of  the expected change in $X_i$ when $r_t(i)< n$ in  \eqref{2.1}.  We see from \eqref{8.1} that
\begin{align*}
d\X(t)&= \sumi dX_i(t)-ngX_{(n)}(t)\,dt\\
&=\sumi \bigg(-g+\frac{\s_{r_t(i)}^2}{2}\bigg)X_i(t)\,dt+\sumi \s_{r_t(i)} X_i(t)\,dW_i(t),\as
\end{align*}
Hence, condition \eqref{3} implies that
\begin{equation}\label{8.2}
\sum_{i=1}^n \bigg(\frac{\s_{r_t(i)}^2}{2g}-1\bigg)X_i(t)=\sum_{k=1}^n \bigg(\frac{\s_k^2}{2g}-1\bigg)X_{(k)}(t)=0,\as,
 \end{equation}
which is a natural generalization of \eqref{2.2}. Since $X(t)>0$, a.s., we can divide by it in \eqref{8.2} and take the expectation, which gives us the condition
\begin{equation}\label{3.1}
\sum_{k=1}^n \bigg(\frac{\s_k^2}{2g}-1\bigg)\E\big[\th_{(k)}(t)\big]=0,
 \end{equation}
where the  expected weights satisfy $\E\big[\th_{(1)}(t)\big]>\cdots>\E\big[\th_{(n)}(t)\big]>0$,  and $\E\big[\th_{(1)}(t)\big]+\cdots+\E\big[\th_{(n)}(t)\big]=1$. 

\vspace{10pt}
\noindent{\bf Definition 2.} A generalized Atlas model of the form \eqref{8} is {\em weakly Zipfian} if \eqref{3.1} holds.
\vspace{10pt}

Now suppose that a generalized Atlas model is weakly Zipfian and that the variances  $\s^2_1<\cdots<\s^2_n$ are increasing with rank.  In this case the values of $\s^2_k/2g$  for the larger weights $\th_{(k)}(t)$, i.e., for smaller $k$,  will be less than one, and the values of $\s^2_k/2g$ for the smaller weights $\th_{(k)}(t)$, i.e., for larger $k$, will be greater than one.  The same will be true for $(\s_k^2+\s_{k+1}^2)/4g$, so  the distribution curve will be concave, with the slope of the tangent flatter than $-1$ for the higher ranks and steeper than $-1$ for the lower ranks. Indeed, as we noted above, concavity of this nature is consistent with many empirical distribution curves: see, e.g., \citeN{Gabaix:1999}, Figure~I, or \citeN{F:2002}, Figure~5.1.

\vspace{10pt}
\noindent{\large\bf Other Zipfian systems}
\vspace{8pt}

In the models \eqref{1} and \eqref{8} all the ranks share a common reversion rate $g$. However, Atlas models can be further generalized to {\em first-order models}, which are systems of the form
\begin{equation}\label{10}
d\log X_i(t) = g_{r_t(i)}dt+\s_{r_t(i)}dW_i(t),
\end{equation}
for $i=1,\ldots,n$, where $\s_1,\ldots,\s_n$ are positive constants, $r_t(i)$ is the rank of $X_i(t)$,  $W_1,\ldots,W_n$ is an $n$-dimensional Brownian motion, and $g_1,\ldots,g_n$ are constants such that
\begin{equation}\label{11}
g_1+\cdots+g_n=0,\quad\text{ and }\quad g_1+\cdots+g_m<0 \text{ for }  m<n.
\end{equation}
 (see, e.g., \citeN{F:2002} or \shortciteN{BFK:2005}). For these models,  
 \[
\E\big[\log \th_{(k)}(t)-\log \th_{(k+1)}(t)\big]=\frac{\s_k^2+\s_{k+1}^2}{-4(g_1+\cdots+g_k)},
\]
for $k=1,\ldots,n-1$, so the tangent to the distribution curve between rank $k$ and rank $k+1$ will have a log-log slope of 
\begin{equation}\label{12}
\frac{\E\big[\log\th_{(k)}(t)-\log\th_{(k+1)}(t)\big]}{\log k-\log(k+1)}=\frac{\s_k^2+\s_{k+1}^2}{-4(g_1+\cdots+g_k)(\log k -\log(k+1))}\cong\frac{k(\s_k^2+\s_{k+1}^2)}{4(g_1+\cdots+g_k)}.
\end{equation}
Note that this slope is consistent with the slopes \eqref{1.21} and \eqref{8.0} for the previous more restrictive models \eqref{1} and \eqref{8}. From \eqref{12} we see that the generalized model \eqref{10} can be parameterized to fit an arbitrary strictly decreasing distribution curve. Nevertheless, these results do not appear to suggest any obvious further generalizations  (at least not to the authors). 
 
There is a further generalization of these models that might be worthy of mention since it would accommodate a generalization of Zipf's law that appears in \citeN{Gabaix:1999}, Section~III.2. In this more general case, we would consider a version of \eqref{10} where the $X_i$ depend on parameters based on both rank and {\em index} or {\em name}. These {\em hybrid Atlas models} were introduced by \shortciteN{IPBKF:2011}, however, parameter estimation for these models has not been completely resolved (as far as the authors know; see \shortciteN{FIK:2012}). 

\vspace{10pt}
\noindent{\large\bf Example: the ``size effect'' for stocks}
\vspace{8pt}

It was observed by \citeN{banz:1981} that stocks of U.S. companies with smaller capitalizations can be expected to have higher returns on average than stocks of U.S. companies with larger capitalizations. The explanation for this ``anomaly'' was considered to be the higher risk involved in holding smaller stocks (see \citeN{Fama/French:1993}). Here we present a simple structural explanation based on the weak version of Zipf's law.

Suppose that the processes $X_i$ in \eqref{8} represent the capitalizations of U.S. companies.  It follows from \eqref{8.1} that the {\em  relative return} of the stock $X_i$ at time $t$ is
\begin{equation}\label{13}
\frac{dX_i(t)}{X_i(t)}=\bigg(-g+\frac{\s^2_{r_t(i)}}{2}+ng\1_{\{r_t(i)=n\}}\bigg)dt+\s_{r_t(i)} dW_i(t),\as,
\end{equation}
and similarly for the more restrictive configuration  \eqref{2}, where $\s$ replaces $\s_{r_t(i)}$. Note that we are using relative return, since we removed the overall growth $\g$ from the general model \eqref{0}. For simplicity, we have ignored the payment of dividends or other distributions as a source of return since it was shown in \citeN{f:cds} that the difference in these payments between large and small U.S. stocks had little influence on the observed size effect.

For a Zipfian model we have $\s_{r_t(i)}^2/2g=\s^2/2g=1$, so it follows from \eqref{13} that
\begin{equation}\label{14}
\E\bigg[\frac{dX_i(t)}{X_i(t)}\bigg|r_t(i)=k\bigg]=\bigg(\frac{\s^2}{2g}-1\bigg)g\,dt=0,\as,
\end{equation}
for $k=1,\ldots,n-1$. Hence, the expected  relative return of each stock above the bottom rank is zero. However, for a Zipfian model the distribution curve will be linear, and we know that the distribution curve for stock capitalizations is concave rather than linear (see, e.g., \citeN{Ijiri/Simon:1974} or \citeN{F:2002}, Figure~5.1). Therefore, the model can be at most weakly Zipfian. 

Suppose  the model is weakly Zipfian with increasing variances $\s^2_1<\cdots<\s^2_n$, which is consistent with a concave distribution curve. Instead of \eqref{14}, we now have
\begin{equation}\label{15}
\E\bigg[\frac{dX_i(t)}{X_i(t)}\bigg| r_t(i)=k\bigg]=\bigg(\frac{\s_k^2}{2g}-1\bigg)g\,dt,\as,
\end{equation}
for $k=1,\ldots,n-1$. Condition \eqref{3.1} and the increasing variances $\s^2_1<\cdots<\s^2_n$ imply that a large stock $X_i$ will have lower variance, so $\s_k^2/2g<1$, and the conditional expectation in \eqref{15} will be negative, while  a small stock $X_i$ will have higher variance, so $\s_k^2/2g>1$, and the conditional expectation in \eqref{15}  will be positive. Hence, the expected return on  small  stocks will be greater than the expected return on  large stocks, and this provides a natural structural explanation for the size effect.

\bibliographystyle{chicago}
\bibliography{math}
\end{document}